# AI Governance in the Context of the EU AI Act: A Bibliometric and Literature Review Approach

**Byeong-Je Kim[1,2], Seunghoo Jeong[3], Bong-Kyung Cho[4], and Ji-Bum Chung[5]**
[1]The Institute of Social Data Science, Pohang University of Science and Technology, Pohang, Republic of Korea
[2]Division of Advanced Nuclear Engineering, Pohang University of Science and Technology, Pohang, Republic of Korea
[3]R&D Planning and Strategy Division, Korea Railroad Research Institute, Uiwang, Republic of Korea
[4]Strategy Group, INTERX, Seoul, Republic of Korea
[5]Department of Civil, Urban, Earth, and Environmental Engineering, Ulsan National Institute of Science and Technology, Ulsan, Republic of Korea

Corresponding author: Ji-Bum Chung (learning@unist.ac.kr)

This work was supported by Korea Institute of Public Administration (KIPA) Research Funding.

**ABSTRACT** The rapid advancement of artificial intelligence (AI) has brought about significant societal changes, necessitating robust AI governance frameworks. This study analyzed the research trends in AI governance within the framework of the EU AI Act. This study conducted a bibliometric analysis to examine the publications indexed in the Web of Science database. Our findings reveal that research on AI governance, particularly concerning AI systems regulated by the EU AI Act, remains relatively limited compared to the broader AI research landscape. Nonetheless, a growing interdisciplinary interest in AI governance is evident, with notable contributions from multi-disciplinary journals and open-access publications. Dominant research themes include ethical considerations, privacy concerns, and the growing impact of generative AI, such as ChatGPT. Notably, education, healthcare, and worker management are prominent application domains. Keyword network analysis highlights education, ethics, and ChatGPT as central keywords, underscoring the importance of these areas in current AI governance research. Subsequently, a comprehensive literature review was undertaken based on the bibliometric analysis findings to identify research trends, challenges, and insights within the categories of the EU AI Act. The findings provide valuable insights for researchers and policymakers, informing future research directions and contributing to developing comprehensive AI governance frameworks beyond the EU AI Act.

## I. INTRODUCTION

Advancements in artificial intelligence (AI) technology have progressed at an unprecedented pace. The proliferation of generative AI models, such as ChatGPT, is accelerating transformations in daily life. Public interest initially centered on industrial capabilities, such as efficient computation and process optimization. Generative AI has gained prominence, sparking both excitement and worries about its effects on society [1].

While AI offers numerous benefits, including increased industrial efficiency and enhanced convenience, it also poses novel risks, such as privacy violations and ethical dilemmas [2]. Given the extensive influence and reach of AI, experts increasingly warn that its unregulated proliferation, absent adequate societal oversight, could precipitate irreversible societal disruption [2, 3].

According to Maslej et al. [4], the growing number of AI misuse and abuse cases worldwide underscores the escalating risks that AI poses to society. These risks encompass a range of concerns, including bias and discrimination, security breaches, privacy violations, and overreliance [2, 3, 5]. Therefore, a balanced approach that integrates policy and societal discussions alongside technological development is essential. Consequently, the need for AI governance, which involves the development of technical, policy, and social solutions to address these risks, is rapidly increasing [3]. Mäntymäki et al. [6] have defined AI governance as "a system of rules, practices, processes, and technological tools that are employed to ensure an organization's use of AI technologies aligns with the organization's strategies, objectives, and values; fulfills legal requirements; and meets principles of ethical AI followed by the organization." AI governance, therefore, plays a critical role in minimizing risks and ensuring ethical, legal, and social responsibility throughout the AI system lifecycle, from development to deployment and use.

The rapid evolution of AI has dramatically amplified its impact across all societal sectors, fueling the demand for robust ethical AI standards and regulatory frameworks. Leading nations in AI technology, including the United States, Canada, and the United Kingdom, are intensifying their efforts to establish regulations for AI's safe development and application. In the United States, the introduction of the Digital Equity Act (2020) and Blueprint for an AI Bill of Rights (2023) represent significant steps toward building a framework for trustworthy and secure AI systems. The European Union (EU) has adopted a proactive approach by proposing the EU AI Act in April 2021, a legislative framework aimed at guiding technological advancements to mitigate the potential societal implications of AI. Finalized in the second quarter of 2024, the Act's core principle is the restriction or prohibition of AI systems that pose risks to human rights and ethical standards.

Despite the implementation of policies for the development of safe and reliable AI systems, academic research on AI has predominantly concentrated on technological development and model optimization [3]. The technology-centric nature of the field has resulted in relatively limited researcher interest in safe and ethical AI. This imbalance could widen the gap between technological advancements and governance, potentially leading to a misalignment between technology and societal norms.

To foster responsible and ethical AI development, research into AI policy is urgently needed. However, most studies on its research trends still focus on technical aspects, with insufficient consideration for governance and policy implications. This study aims to bridge the gap between AI technology and social norms by conducting a bibliometric analysis combining a literature review of AI governance research performed within the framework of the EU AI Act. The findings will help researchers in various fields understand the trends and key issues in AI governance research, providing a foundation for further research on responsible AI development.

## II. THE EU AI ACT

The EU AI Act aims to ensure the trustworthy development and deployment of AI systems for EU citizens while fostering a robust AI ecosystem. Moreover, the Act strives to protect individual safety and fundamental rights, strengthen transparency and accountability, promote innovation, and maintain a stable market environment. The EU AI Act defines an AI system as a machine-based system designed to operate with varying degrees of autonomy and adaptability, and produces outputs such as predictions, content, recommendations, or decisions, influencing physical or virtual environments. The Act adopts a risk-based regulatory framework, categorizing AI systems based on their potential risks to health, safety, and fundamental rights. AI systems with unacceptable risks are classified as "prohibited AI systems," while systems with higher but acceptable levels of risk are categorized as "high-risk AI systems." These two categories form the primary focus of the regulation. Table I summarizes the types of AI systems classified as prohibited or high-risk under the EU AI Act and provides descriptions of each.

The prohibited systems, which are subject to the strictest regulation, include those that deploy subliminal techniques to manipulate human behavior, exploit human vulnerabilities, or violate fundamental rights. These systems contravene the EU's core values of human dignity, freedom, equality, non-discrimination, democracy, and respect for the rule of law. Examples of prohibited systems include AI systems that manipulate individuals through subliminal techniques beyond their conscious awareness, systems exploiting specific vulnerabilities of individuals or groups (e.g., age, disability, or socio-economic status), general-purpose social scoring systems (e.g., systems analogous to China's Social Credit System), and real-time remote biometric identification systems in publicly accessible spaces. Additionally, systems that infer sensitive information from biometric data, predict criminal risks, indiscriminately collect biometric data from non-targeted individuals, or automatically recognize the emotions of workers or students are included in the prohibited category.

TABLE I
CLASSIFICATION AND DESCRIPTION OF PROHIBITED AND HIGH-RISK AI SYSTEMS UNDER THE EU AI ACT

| AI system | Type | Description |
| --- | --- | --- |
| Prohibited AI Systems | P1 | Subliminal manipulation |
| | P2 | AI systems that exploit people's vulnerabilities |
| | P3 | AI systems that evaluate or classify people based on their social behavior or personal traits |
| | P4 | AI systems that predict a person's risk of committing a crime |
| | P5 | Untargeted scraping of facial images from the internet or CCTV footage |
| | P6 | AI systems that infer emotions in the workplace or educational institutions |
| | P7 | AI systems that categorize people based on their biometric data |
| | P8 | Real-time remote biometric identification in publicly accessible spaces |
| High-risk AI Systems | H1 | Biometric identification and categorization |
| | H2 | Critical infrastructure management |
| | H3 | Educational and vocational training |
| | H4 | Employment, worker management, and access to self-employment |
| | H5 | Access to essential private and public services |
| | H6 | Law enforcement |
| | H7 | Border control and migration management |
| | H8 | Administration of justice and democratic processes |

The EU AI Act clearly distinguishes between the concepts of Foundation Models and General-Purpose AI Systems. Following the so-called "ChatGPT shock," the regulation of foundation models emerged as a priority. However, the EU chose to regulate not the models themselves but the risks associated with the AI systems and applications that integrate such models. Accordingly, the EU AI Act explicitly incorporates foundation models within its risk-based framework, focusing on assessing and regulating risks at the system level rather than targeting the models directly. This approach provides a theoretical basis for this study's analysis of governance trends at the AI system level.

Globally, there is also a growing movement toward AI regulation. In October 2023, the United States issued an executive order aimed at the development and risk management of AI systems, with state-level regulations under active discussion. In California, home to leading AI innovators, legislative proposals targeting AI developers are being actively debated. In China, the government has implemented direct and strict regulations, including algorithm registration and review systems, AI content censorship, and data verification. Japan, on the other hand, announced AI Governance Guidelines for Business Version 1.0 (April 2024) and has established an organization named "AI Safety" to review evaluation and risk assessment methods for ensuring AI safety. South Korea, on December 26, 2024, passed a comprehensive AI legislative framework through its National Assembly plenary session, becoming the second in the world after the EU to establish such an overarching AI regulatory act.

Based on the aforementioned background, this research aims to analyze the trends in academic research related to AI governance. Specifically, the study will focus on the research trends for core AI systems regulated by the EU AI Act. To achieve this purpose, the following research questions are posed:

1) What is the extent of progress made in research on both overall AI governance and governance of risky AI systems covered by the EU AI Act?

2) Which countries are leading in researching AI governance related to the EU AI Act?

3) Through text mining, what are the predominant research topics in the field of AI governance?

This study seeks to contribute by providing answers to these questions and gaining insights into the current state and challenges of AI governance research, as well as identifying potential future research directions.

## III. METHODOLOGY

Bibliometric analysis offers a data-driven approach to comprehensively understand a research field. It is effective for exploring interdisciplinary and convergent research areas [7, 8]. Quantitatively assessing the metadata of publications, allows for the systematic organization of core concepts and research methods embedded within a large number of academic publications. It also facilitates the effective sharing of synthesized scientific knowledge among researchers. The sharing of systematically organized and analyzed bibliographic metadata inspires other researchers, serving as a driving force for knowledge dissemination [9-11].

This study employed text mining-based keyword analyses to examine which research topics have been assessed in the context of AI governance research. The research trends of AI governance studies relevant to the EU AI Act were analyzed based on the research framework shown in Fig. 1. This study consists of: (1) data collection, (2) data screening, (3) data preprocessing, (4) text mining, and (5) understanding of the research trends.

Subsequently, a comprehensive literature review was undertaken based on the bibliometric analysis findings to identify research trends, challenges, and insights within the categories of the EU AI Act.

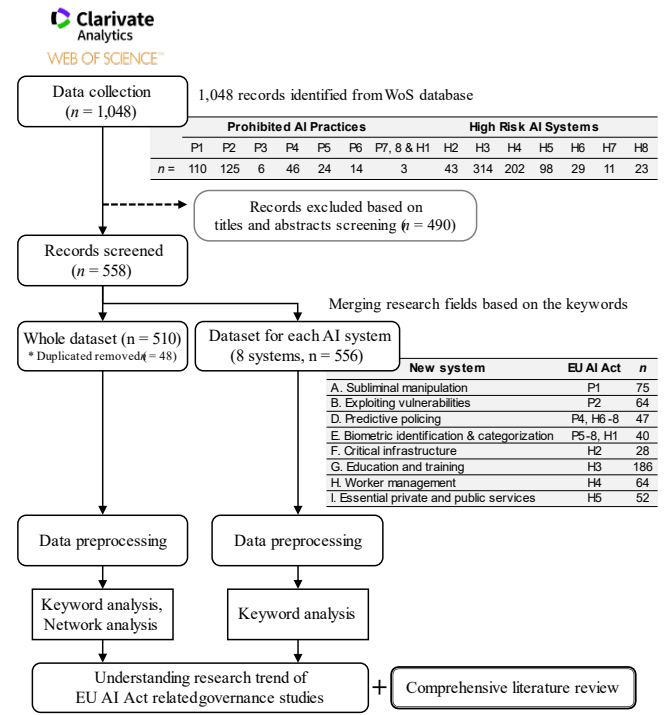

**FIGURE 1. Research flow summary**

### A. DATA COLLECTION AND SCREENING

Relevant publications were collected from the Web of Science (WOS) Core Collection database, utilizing the "topic" search option. The "Topic" field in the Web of Science encompasses the title, abstract, author keywords, and Keywords Plus. The search was conducted on October 11, 2024, with no restrictions placed on the publication period.

Search keywords were structured as "(AI OR "artificial intelligence") AND (regulation OR governance) AND *system-specific keywords*" to ensure the retrieval of studies related to AI governance. These keywords were established by referring to prior research and the provisions of the EU AI Act [5, 12]. Upon reviewing the search keywords for the 16 types of AI systems restricted by the AI Act, it was found that the keywords for P7 and P8 of the Prohibited AI Practices, and H1 of the High-Risk AI systems were identical. Consequently, this study consolidated similar AI systems into nine distinct categories. The research trends for each redefined system were then searched. The specific search keywords for each AI system are detailed in Table II.

While the initial search aimed to include studies highly relevant to AI governance based on the keywords, the collected data might include studies with varying degrees of relevance. Therefore, a screening process was conducted to exclude studies with low relevance. The authors assessed the relevance of the retrieved publications to relevant AI systems based on titles and abstracts. Following this process, the final papers for analysis were selected.

TABLE II
WOS SEARCH KEYWORDS FOR REDEFINED AI SYSTEMS

| Redefined AI System | EU Definition | Search keywords |
|---|---|---|
| A. Subliminal manipulation | P1 | (subliminal OR manipul* OR decept*) |
| B. Exploiting vulnerabilities | P2 | (child* OR disab* OR elder*) |
| C. Social scoring | P3 | ("social behavior" OR "personal trait") |
| D. Predictive policing | P4 | (crime OR criminal) |
|  | H6 | ("law enforcement") |
|  | H7 | ("border control" OR visa OR asylum OR immigra* OR emigra*) |
|  | H8 | ("judicial authority" OR election OR referendum OR vote) |
| E. Biometric identification and categorization | P5 | ("facial recognition") |
|  | P6 | ("emotion infer*" OR "emotion recog*" OR "emotion identif*") |
|  | P7 P8 H1 | ("biometric categor*" OR "biometric identif*") |
| F. Critical infrastructure | H2 | (critical infrastructure) |
| G. Education and training | H3 | ("education" OR "vocational") |
| H. Worker management | H4 | (employment OR employee* OR worker* OR recruit*) |
| I. Essential private and public services | H5 | ("private services" OR "public services" OR "essential services" OR "triage" OR "credit score" OR creditworthiness OR insurance) |

### B. DATA PREPROCESSING

The present work performed data preprocessing to ensure that the collected bibliographic information was appropriate for text analysis. Data preprocessing was performed using the R programming language. This process included lemmatization and the removal of repetitive phrases and stopwords. Text refinement and analysis were conducted on the titles, abstracts, and keywords, respectively. Each word was lemmatized to its standardized form using the 'textstem' package in R. This process ensures that variations of a word, such as plurals or different verb tenses, are treated as a single term, thereby enhancing the efficiency of the analysis. Repetitive phrases, such as copyright notices from publishers, were also removed. Furthermore, a list of stopwords was added to the basic stopwords list in the 'tm' package to eliminate terms related to AI technologies and models (Table III). As this study does not focus on AI research tools and techniques, additional stopwords were incorporated based on prevalent AI research tools and techniques found in existing studies [1, 5].

TABLE III
LIST OF ADDITIONAL STOPWORDS

| 'technology', 'good', 'new', 'analysis', 'internet of thing', 'systematic review', 'literature review', 'scope review', 'large language model', 'natural language', 'machine learn', 'neural network', 'deep learn', 'federate learn', 'reinforcement learn', 'learn', 'model', 'ai', 'artificial', 'intelligence', 'differential', 'equation', 'dynamic', 'datum', 'system', 'use', 'can', 'result', 'show', 'train', 'novel', 'algorithm', 'linear', 'graph', 'ann', 'network', 'deep', 'code', 'real', 'world', 'e g', 'u s', 'et al', 'state', 'art', 'time', 'large scale', 'open source', 'fine tune', 'long term', 'high', 'task', 'dataset', 'recent', 'github', 'http', 'https', 'com', 'url', 'support vector machine', 'due', 'play', 'role', 'make', 'base', 'much', 'one', 'two', 'work', 'may', 'demonstrate', 'need', 'will', 'first', 'many', 'far', 'within', 'aim', 'end', 'predict', 'optimization', 'via', 'towards', 'propose', 'process', 'approach', 'method', 'provide', 'paper', 'research', 'study', 'also', 'present', 'however', 'include', 'aaai fss', 'proceeding', 'challenge' |
|---|

### C. TEXT MINING

For descriptive analysis, statistics on the annual publication trends and the countries of corresponding authors were compiled. Text analysis was performed by calculating keyword frequency. Furthermore, to determine the context of key terms, network analysis was conducted on the author-provided keywords. Network analysis is a bibliometric method that interprets the connections and interactions among various actors in a network structure, analyzing relationships formed by interactions within complex systems [13]. This study employed Gephi, an open-source network analysis software, to perform the network analysis [14].

A network consists of nodes and edges (links). In the context of author keyword networks, nodes represent keywords used in the studies, and edges represent the co-occurrence frequency of keywords within the same study. The edge weight is proportional to the number of co-occurrences between keywords. For instance, if two keywords appear together in one paper, an edge weight of one is assigned. The network was designed as undirected, assuming no directionality between keywords. The resulting network was analyzed using centrality measures and a community detection algorithm.

Centrality is a widely used method to assess the importance of a node relative to others in the network [15]. Among various centrality measures, this study focused on degree centrality to gauge the overall research interest in specific topics. Degree centrality indicates the number of connections a node has with other nodes. A higher degree centrality value suggests that the node appears more frequently with other nodes. Analyzing centrality values helps identify which research topics are considered important in the network. Important keywords were identified based on centrality metrics. By examining the edge weights, which signify the co-occurrence frequency between keywords, the main research themes and the relationships between them in AI governance research were understood.

Community clusters within the keyword network were identified using modularity analysis, a technique that measures the strength of network division into modules [16, 17]. Modularity analysis was performed in Gephi, allowing for the interpretation of each cluster and their interrelationships based on their constituent keywords.

## IV. RESULTS
### A. DESCRIPTIVE ANALYSIS

As of October 2024, a search of the Web of Science database using the keywords ("AI" OR "artificial intelligence") yielded over 220,000 studies. When the search was refined to "(AI OR "artificial intelligence") AND (governance OR regulation)," approximately 6,000 studies were identified.

A search for research trends related to the 16 AI systems defined in the EU AI Act identified 1,048 studies. After a review of titles and abstracts, 490 studies with low relevance were excluded. This process resulted in 558 studies related to governance and regulation relevant to the EU AI Act. After removing duplicates between each system, 510 unique documents were identified. This represents approximately 7% of the research related to AI governance and regulation. The 510 unique publications were analyzed for the overall research trend analysis and network analysis. To understand trends in each field, analyses were conducted separately for each AI system.

The number of publications for each system is presented in Table IV. The field with the most relevant literature was Type G (186 studies), which involves the use of AI in education or vocational training and learner assessment. This was followed by Type A systems (86 studies), which influence individual choices through AI. Type B systems, which target vulnerable populations, and Type H systems, which are related to worker management, also featured prominently, with 64 studies each.

Despite the recent surge in interest in AI-related research, search results show that studies on Type C system were relatively scarce. Only two studies on this system were identified in this research. Due to this limited number of documents, the authors determined that a keyword-based bibliometric analysis would not be effective for this category. Thus, for the system-specific analysis, research trends were examined using 556 studies across the eight remaining AI systems.

TABLE IV
NUMBER OF PUBLICATIONS FOR EACH AI SYSTEM AFTER SCREENING

| Redefined AI System | AI System (EU AI Act) | Records |
|---|---|---|
| A. Subliminal manipulation | P1 | 75 |
| B. Exploiting vulnerabilities | P2 | 64 |
| C. Social scoring | P3 | 2 |
| D. Predictive policing | P4, H6, H7, H8 | 47 |
| E. Biometric identification and categorization | P5, P6, P7, P8, H1 | 40 |
| F. Critical infrastructure | H2 | 28 |
| G. Education and training | H3 | 186 |
| H. Worker management | H4 | 64 |
| I. Essential private and public services | H5 | 52 |

AI governance-related literature began to be published in significant numbers from 2017 onwards, with a notable surge in the number of related studies in the 2020s (Fig. 2). Despite the search results for 2024 spanning only from January to October, the number of published studies in most fields has already surpassed that of the previous year. Types A, B, G, and H systems, in particular, have shown a consistent trend of annual increases in the number of published studies. Notably, Type B systems saw a significant jump in publications, with 28 papers published up to October 2024, compared to only 8 in the previous year. Similarly, for Type G systems, the number of papers published more than doubled year over year, from 37 to 95.

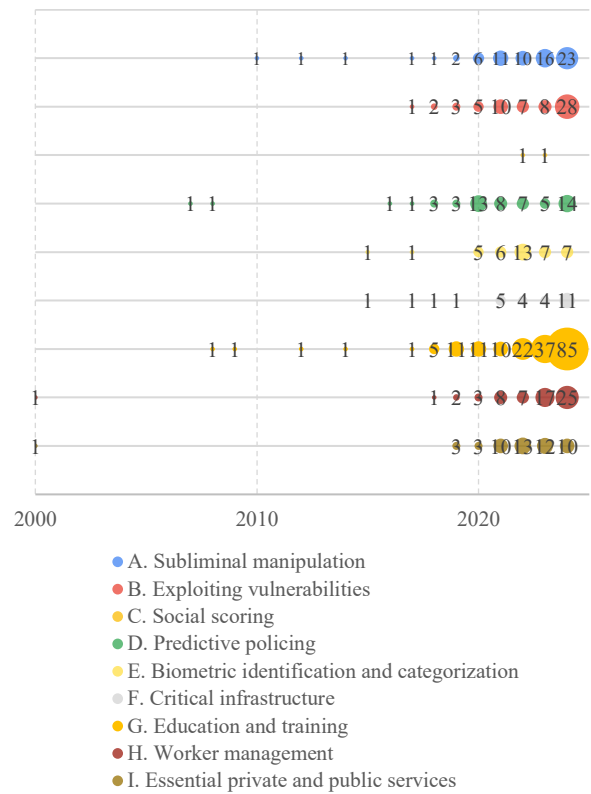

FIGURE 2. Number of Annual Publications by AI Systems

Table V presents the journals and publishers with the highest number of publications related to this research. *Sustainability* published the most studies, with a total of 14. *Computer Law & Security Review* (12 articles), *IEEE Access* (7 articles), and *Government Information Quarterly* (7 articles) also demonstrated significant interest in AI governance research. These results indicate that research on AI governance is being conducted across various disciplines, including not only technical journals but also multidisciplinary journals, policy-focused journals, and journals specializing in education and healthcare. Notably, the proportion of open-access journals and publishers was particularly high.

TABLE V
TOP JOURNALS AND PUBLISHERS PUBLISHING RELEVANT RESEARCH.

| Journal (Publisher) | Number of articles |
|---|---|
| Sustainability (MDPI) | 14 |
| Computer law & security review (Elsevier) | 12 |
| IEEE Access (IEEE) | 7 |
| Government information quarterly (Elsevier) | 7 |
| British Journal of Educational Technology (Wiley) | 6 |
| Frontiers in Public Health (Frontiers) | 6 |
| Learning Media and Technology (Taylor & Francis) | 6 |
| Technology in Society (Wiley) | 5 |
| Applied Sciences-Basel (MDPI) | 4 |
| Education and Information Technologies (Springer) | 4 |
| International Journal of Educational Technology in Higher Education (Springer) | 4 |
| Journal of Computer Assisted Learning (Wiley) | 4 |
| Journal of Medical Internet Research (JMIR) | 4 |
| Policy And Internet (Elsevier) | 4 |

Table VI displays the number of publications per country based on the corresponding author's affiliation. The analysis of publication counts across countries reveals variations in research directions and approaches to AI regulation and governance, contingent upon specific national contexts.

The analysis of the number of publications by country generally shows that AI governance research is most active in the United States, followed by China, the United Kingdom, and Australia. The U.S. has published the most publications in eight of the nine AI system categories, the exception being Type G. China shared the top position with the U.S. for Type B and E systems and published the most studies in Type G systems. The U.K. is one of the countries with a high level of interest in AI research in general. It tied with the U.S. for the highest number of publications on Type D systems and had the third-highest number of publications on Type B, H, and I systems, after the U.S. and China. For Type G systems, the order of publication count was China, the U.S., Australia, and the U.K., which is consistent with the findings of previous research on AI in education [18]. Regarding Type I systems, the U.S. published seven studies, while China published only two, indicating a difference in the level of interest. In the case of Type I systems, several studies were also published in Australia and the U.K., in addition to the U.S.

TABLE VI
NUMBER OF PUBLICATIONS PER COUNTRY BASED ON THE CORRESPONDING AUTHOR'S AFFILIATION

| AI System | Top publication countries (n. of publications) | Records |
|---|---|---|
| A. Subliminal manipulation | **USA (11), China (9),** Italy (5), England, Germany, South Korea (4), Malaysia, Netherlands, Scotland, Spain, Wales (3), Australia, Canada, Israel, Japan, Russia (2) | 75 |
| B. Exploiting vulnerabilities | **China (13), USA (13),** England (8), Greece (4), Canada, Italy, Netherlands, Norway, Portugal, Singapore, Sweden, Wales (2) | 64 |
| C. Social scoring | Germany, USA (1) | 2 |
| D. Predictive policing | **England, USA (7),** China, Netherlands (4), Russia, Spain (3), Germany, South Korea (2) | 47 |
| E. Biometric identification and categorization | **China, USA (7),** Belgium, Canada, England, Netherlands, Portugal (2) | 40 |
| F. Critical infrastructure | **USA (5),** China, India (3), Italy (2) | 28 |
| G. Education and training | **China (35), USA (32),** Australia, England (13), Germany (9), Canada, Italy, South Africa, Sweden (5), Greece, India, Scotland, Spain, Switzerland (4) | 186 |
| H. Worker management | **USA (10), China (7),** England, Spain (4), Australia, Poland, Taiwan (3), Germany, Romania, Serbia, South Africa, Ukraine (2) | 64 |
| I. Essential private and public services | **USA (7), Australia, England (6),** India, Italy, Taiwan (3), China, Estonia, Finland, France, Germany, Netherlands, Spain (2) | 52 |
| Total | **USA (84), China (76), England (40),** Australia (27), Germany (20), Italy (18), Spain (15), Netherlands (14), Canada (12), India (12), Taiwan (9) | 510 |

### B. RESEARCH TRENDS OF AI SYSTEMS
#### 1) KEYWORD ANALYSIS

To identify the primary areas of interest in the relevant research, this study analyzed frequently occurring keywords. Table VII presents the list of the most frequent keywords in titles, abstracts, and author-provided keywords after the removal of stopwords. As revealed in the search results of the relevant literature, studies related to education were prevalent across titles, abstracts, and keywords. Key AI governance-related keywords identified include *risk*, *ethic(al)*, *policy*, and *privacy*.

The prominence of terms such as *generative AI* and *ChatGPT* indicates that alongside the expanding application scope of AI systems, there is a growing body of research focused on the ethical considerations and potential side effects of their deployment. In particular, the keyword analysis highlighted a significant interest in AI ethics, alongside numerous education-related keywords. Furthermore, frequent mentions of the EU's General Data Protection Regulation (*GDPR*) and related terms, as well as *transparency*, were observed in relation to potential privacy infringements associated with AI technology use.

TABLE VII
MOST FREQUENT WORDS IN TITLES, ABSTRACTS, AND KEYWORDS OF AI GOVERNANCE STUDIES RELATED TO THE EU AI ACT

| Rank | Title Word | Freq | Abstract Word | Freq | Author keywords Word | Freq |
|---|---|---|---|---|---|---|
| 1 | education | 72 | education | 358 | ethic | 27 |
| 2 | future | 29 | development | 276 | education | 26 |
| 3 | public | 29 | human | 265 | chatgpt | 23 |
| 4 | human | 28 | application | 234 | big_data | 14 |
| 5 | application | 26 | potential | 233 | generative_ai | 13 |
| 6 | health | 26 | digital | 230 | chatbot | 13 |
| 7 | digital | 25 | public | 221 | high_education | 13 |
| 8 | framework | 24 | risk | 210 | privacy | 13 |
| 9 | healthcare | 21 | service | 208 | data_protection | 12 |
| 10 | ethical | 20 | ethical | 206 | gdpr | 9 |
| 11 | service | 19 | health | 204 | transparency | 9 |
| 12 | generative | 18 | policy | 204 | learn_analytics | 8 |

| 13 | policy | 18 | framework | 201 | self-regulate_learn | 8 |
| 14 | social | 18 | design | 197 | self-regulation | 8 |
| 15 | chatgpt | 17 | healthcare | 196 | child | 7 |
| 16 | design | 17 | decision | 194 | explainable_ai | 7 |
| 17 | self | 17 | develop | 187 | innovation | 7 |
| 18 | law | 16 | tool | 181 | public_service | 7 |

Table VIII presents the top 10 author keywords for each system type, excluding common keywords (e.g., *AI, regulation, governance*) and system-specific keywords (e.g., *insurance, education, employee*) used for the search.

TABLE VIII
FREQUENTLY USED TERMS IN ABSTRACT BY AI SYSTEMS

| | System A | | System B | | System D | | System E | |
|---|---|---|---|---|---|---|---|---|
| | Word | Freq | Word | Freq | Word | Freq | Word | Freq |
| 1 | human | 65 | care | 60 | legal | 50 | ethical | 26 |
| 2 | market | 41 | social | 54 | risk | 36 | privacy | 23 |
| 3 | development | 38 | design | 47 | framework | 31 | social | 23 |
| 4 | activity | 35 | development | 44 | digital | 30 | human | 21 |
| 5 | ethical | 34 | healthcare | 41 | public | 28 | information | 21 |
| 6 | potential | 34 | health | 38 | human | 27 | development | 20 |
| 7 | design | 33 | human | 38 | decision | 26 | law | 19 |
| 8 | application | 31 | policy | 37 | policy | 26 | service | 19 |
| 9 | image | 31 | risk | 35 | social | 26 | application | 18 |
| 10 | control | 30 | develop | 34 | development | 24 | right | 18 |

| | System F | | System G | | System H | | System I | |
|---|---|---|---|---|---|---|---|---|
| | Word | Freq | Word | Freq | Word | Freq | Word | Freq |
| 1 | digital | 31 | student | 175 | human | 57 | health | 50 |
| 2 | service | 25 | chatgpt | 129 | digital | 48 | government | 43 |
| 3 | solution | 21 | development | 108 | management | 48 | healthcare | 43 |
| 4 | control | 20 | application | 106 | health | 35 | patient | 40 |
| 5 | transformation | 20 | potential | 106 | potential | 32 | sector | 37 |
| 6 | public | 19 | healthcare | 95 | safety | 32 | risk | 34 |
| 7 | quality | 19 | ethical | 89 | job | 31 | information | 28 |
| 8 | information | 18 | support | 89 | labor | 30 | policy | 28 |
| 9 | integrate | 17 | practice | 87 | decision | 28 | enable | 27 |
| 10 | decision | 16 | tool | 83 | framework | 28 | framework | 27 |

System A: Subliminal manipulation, B: Exploiting vulnerabilities, D: Predictive policing, E: Biometric identification and categorization, F: Critical infrastructure, G: Education and training, H: Worker management, I: Essential private and public services.

In Type A systems (Subliminal Manipulation), keywords such as *human, market, design*, and *image* appeared frequently. This suggests active research into marketing applications of this technology and image-focused techniques for influencing the subconscious. The frequent appearance of the keyword *ethical* also indicates that the ethical considerations of this technology are being raised.

For Type B systems (Exploiting Vulnerabilities), keywords such as *healthcare, social*, and *design* appeared frequently, reflecting a high level of interest in medical technologies for vulnerable populations, including the elderly and children. Although not shown in Table VIII due to the removal of the search terms, the primary subjects of research in this area were children (appearing 84 times) and the elderly (appearing 31 times). The main research topics also included AI-based education methods for children and personalized services for vulnerable populations. The keyword *risk* reflects concerns about the potential dangers and safety of AI systems targeting vulnerable groups.

In Type D systems (Predictive Policing), keywords such as *legal, risk, framework, digital*, and *policy* were frequently used. This relates to active discussions on the ethical and legal risks, including decision-making and legal accountability when using AI in law enforcement and judicial systems. It also reflects scholarly interest in the digital transformation of the judicial system, policy changes, and societal impacts.

For Type E systems (Biometric Recognition), keywords such as *ethical* and *privacy* were predominantly featured. This indicates ongoing discussions regarding potential privacy violations, ethical issues, and legal regulations related to the use of biometric recognition technologies. The frequent appearance of keywords like *social* and *human* suggests that emotion recognition technology is mainly used for analyzing social interactions.

In Type F systems (Critical Infrastructure), keywords such as *digital, public*, and *service* appeared frequently, indicating that this system category includes research related to digital transformation such as smart cities, public service improvement, and infrastructure management. Research has primarily focused on the application of AI systems in public policy, public services, and urban infrastructure.

In Type G systems (Education), keywords such as *student, chatgpt*, and *tool* appeared frequently. This suggests that AI systems in this field are mainly used in AI-based educational tools, learner assessment, and personalized learning support. Notably, the keyword *healthcare* appeared frequently, demonstrating significant interest in the application of AI technology in medical education. The keyword *ethical* also appeared frequently.

In Type H systems (Worker Management), keywords such as *human, digital, health, safety*, and *management* ranked highly, indicating active research on AI-based management systems for worker safety, health management, and work efficiency improvement.

Finally, in Type I systems (Essential Private and Public Services), keywords related to healthcare, such as *health, healthcare*, and *patient*, appeared frequently. The frequent appearance of these healthcare-related terms indicates a high level of interest in the application of AI in essential public services, particularly in healthcare and insurance. It also suggests that patient privacy protection is a major issue in the deployment of such AI systems.

2) NETWORK ANALYSIS
To provide a comprehensive view of the knowledge structure across the AI governance research, a network analysis based on author keywords was conducted (Fig. 3). Due to the insufficient number of studies in each specific field, publications from all system types were combined for analysis. The co-occurrence analysis of author keywords reveals that AI ethics and privacy concerns related to the use of AI technology, along with research in the field of education, are major areas of interest in the related research. Analyzing 510 publications (excluding duplicates between systems) identified 4,730 co-occurrence relationships among 1,478 keywords. The network was visualized based on relationships between keywords that co-occurred at least twice, to simplify the network structure and highlight the most important keywords. As a result, a network of 75 keywords and 132 connections were derived, with a maximum co-occurrence frequency of five.

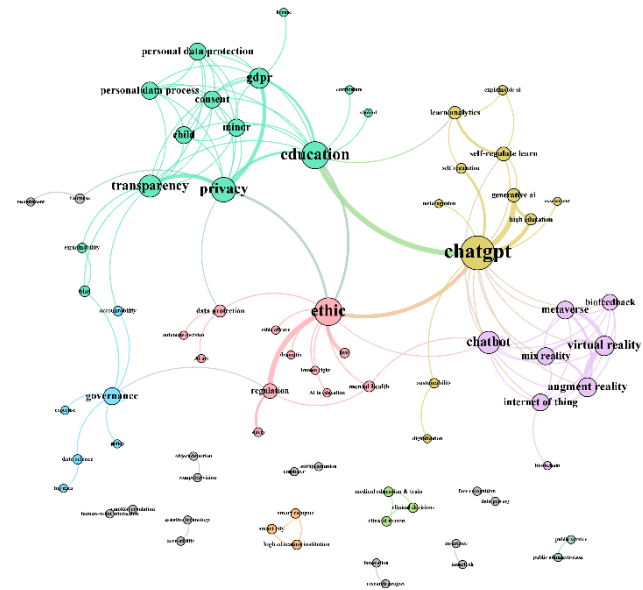

**FIGURE 3.** Keywords Network of AI governance publications

In the network, the size of the nodes was set to be proportional to their occurrence frequency, that is, 'degree centrality.' The thickness of the edges was configured to be proportional to the co-occurrence frequency of the two nodes.

The top keywords based on degree centrality were *education, ethic, ChatGPT, regulation, governance, privacy, chatbot*, and *GDPR* (Table IX). Excluding the keywords used in the search, *ChatGPT, privacy, chatbot, GDPR, big data, transparency*, and *generative AI* exhibited high degree centralities. Notably, *big data* and *ethic* were the keywords with the highest betweenness centrality (0.073 and 0.067, respectively), indicating they are central themes in recent AI governance research.

There are ten pairs that showed a co-occurrence frequency of four or more. The most frequent co-occurring keyword pairs were *AR-VR* (5 times), *ChatGPT-Education* (5), *Generative AI-ChatGPT* (5), *regulation-ethic* (5), *GDPR-privacy* (4), *ChatGPT-ethic* (4), *high education-ChatGPT* (4), *law-ethic* (4), *metaverse-VR* (4), and *transparency-privacy* (4). *ChatGPT* showed high co-occurrence with keywords such as *education, generative AI, ethic*, and *high education*, suggesting active discussions related to the use and regulation of generative AI technology in those fields.

TABLE IX
TOP 20 KEYWORDS BASED ON WEIGHTED DEGREE CENTRALITY (WDC)

| Rank | Keyword | DC | WDC | BC | EC |
|---|---|---|---|---|---|
| 1 | education | 102 | 121 | 0.097 | 1.000 |
| 2 | ethic | 82 | 103 | 0.067 | 0.546 |
| 3 | chatgpt | 74 | 101 | 0.062 | 0.825 |
| 4 | regulation | 91 | 99 | 0.074 | 0.117 |
| 5 | governance | 75 | 83 | 0.064 | 0.083 |
| 6 | privacy | 60 | 77 | 0.027 | 0.893 |
| 7 | chatbot | 64 | 76 | 0.034 | 0.734 |
| 8 | gdpr | 58 | 70 | 0.038 | 0.806 |
| 9 | big data | 63 | 64 | 0.073 | 0.005 |
| 10 | transparency | 46 | 59 | 0.016 | 0.839 |
| 11 | generative ai | 44 | 51 | 0.029 | 0.157 |
| 12 | virtual reality | 34 | 47 | 0.006 | 0.536 |
| 13 | self-regulation | 43 | 46 | 0.023 | 0.130 |
| 14 | child | 38 | 46 | 0.016 | 0.795 |
| 15 | data protection | 40 | 44 | 0.021 | 0.188 |
| 16 | augment reality | 32 | 44 | 0.005 | 0.536 |
| 17 | high education | 35 | 40 | 0.016 | 0.127 |
| 18 | covid-19 | 39 | 39 | 0.021 | 0.000 |
| 19 | internet of thing | 31 | 39 | 0.024 | 0.545 |
| 20 | mental health | 34 | 37 | 0.009 | 0.182 |

DC: Degree centrality, WDC: Weighted degree centrality, BC: Betweenness centrality, EC: Eigenvector centrality.

Community analysis, based on modularity, identified five major clusters. In Fig. 3, each keyword is color-coded according to its cluster. The main clusters were related to *education, ethical*, and *generative AI* (*ChatGPT*), which correspond to the keywords with the highest weighted degree centrality. Additional clusters were identified related to AI-related digital environments, including *chatbot, AR, VR*, and *IoT*, as well as a *governance*-related cluster.

The 'education' cluster was centered around the keyword *education*, the most frequently occurring keyword, and included terms such as *GDPR, privacy, transparency*, and *consent*. This indicates that the main topics of discussion in the application of AI in education include student privacy and the need for informed consent.

In the 'ethics' cluster, keywords used in the search, such as *regulation, data protection*, and *mental health*, appeared. Although the connectivity between the *ethics* and other keywords in this cluster was low, it suggests that data protection, regulatory compliance, and the protection of vulnerable populations are being considered in relation to AI ethics. The connection between the keywords *AI Act* and *data protection* indicates that while direct mentions of the EU AI Act in current research are relatively infrequent, when it is discussed, it is in relation to data protection.

The 'generative AI' cluster, led by *ChatGPT*, included keywords such as *generative AI, higher education*, and *self-regulated learning*, indicating a close relationship with the field of education. This suggests that generative AI is being actively used in education, particularly to support self-regulated learning among higher education students.

The 'digital environment' cluster mainly featured technologies enabling digital environments, such as *chatbot, VR, AR,* and *metaverse*. These keywords were closely linked to *ChatGPT*, indicating they appeared in the context of generative AI implementation.

Finally, the 'governance' cluster was relatively small, despite governance being included as a search term for constructing the dataset. This cluster included keywords such as *policy* and *data science*, suggesting discussions on policies and methodologies for AI governance. Additionally, the *accountability* appeared alongside these terms, which was in turn connected to the *transparency*. This implies that key issues in AI governance include ensuring the transparency and explainability of AI systems and clarifying the accountability for the impacts caused by these systems.

## V. DISCUSSION

This study comprehensively analyzes trends in AI governance research based on the EU AI Act framework. A search of the Web of Science database revealed that studies addressing AI governance and regulation are relatively scarce, particularly in the context of policies related to AI systems regulated by the EU AI Act. This suggests that policy and institutional discussions lag behind the rapid pace of AI technological advancement. Following the risk-based approach outlined in the EU AI Act, the study categorized AI systems by type and analyzed research trends for each category. Findings indicate that interdisciplinary journals and open-access platforms have published a significant number of AI governance studies, highlighting the field's cross-disciplinary nature and the importance of rapid dissemination of research. Furthermore, while the U.S. and China dominate AI governance research, cultural and national backgrounds were found to influence research trends. Text mining and network analysis identified key themes such as ethical considerations, data protection (e.g., GDPR), and generative AI (e.g., ChatGPT), underscoring growing academic interest in ethical issues and privacy concerns. Notably, the prominence of educational applications and the strong network centrality of ChatGPT signal that generative AI is a focal point in recent governance research.

Building upon these general findings, the following subsections present a more granular qualitative literature review of research pertaining to each specific AI system category, extracting valuable insights and identifying future research directions for each.

### A. SUBLIMINAL MANIPULATION AND EXPLOITATION OF VULNERABILITIES

The findings of this study indicate that current research on AI systems focuses primarily on safeguarding vulnerable groups, such as children, seniors, and individuals with disabilities, from subliminal manipulation and exploitation [19-25]. Other areas of concern include ethical considerations for care bots [20, 26], ensuring transparency in consumer decision-making support systems [27], and potential risks associated with conversational AI [28, 29]. The need for stringent regulations for AI systems targeting children [19, 24], ethical considerations in the design of care bots [20, 26], transparency in consumer data-based AI systems [27], and acknowledgment of the potential for misuse and the necessity of regulation for conversational AI [28] have all been emphasized. Furthermore, the possibility of AI systems unintentionally manipulating humans has been raised [30], underscoring the urgent need for sustainable regulatory measures to enhance algorithm transparency [25, 31].

Previous research has predominantly focused on vulnerability exploitation, specifically in vulnerable populations such as children [21, 22, 24], patients, the elderly, and individuals with disabilities [23]. Studies have also explored the implications of deepfake technology [25].

Several key concerns have been raised regarding the potential harms of AI systems. Studies have highlighted the negative impacts of media and AI exposure on children [21], underscoring the need for AI literacy education tailored to young children [24]. In healthcare, research emphasizes the importance of incorporating the perspectives of child and adolescent patients when implementing AI technologies [22]. Further research has stressed the need for ethical guidelines on using intelligent assistive technology for the elderly and individuals with disabilities, particularly addressing issues of autonomy, data management, and distributive justice [23]. The potential for deepfake technology to cause societal harm, especially by facilitating online abuse and violating women's rights, has also been identified as a significant concern [25].

Overall, to minimize the risks of subconscious manipulation and exploitation of vulnerabilities, AI systems must be designed with greater transparency and explainability [25, 31]. Furthermore, it is essential to adhere to principles of personal data protection and non-discrimination. Tailored regulations are also needed to address the specific characteristics of different fields, including protecting vulnerable populations, establishing ethical guidelines for chatbots, safeguarding consumers, and regulating deepfakes. To achieve this, collaboration among various stakeholders, including technology developers, policy makers, ethicists, and citizen groups, is imperative in order to reach a social consensus and establish a sustainable AI governance system.

### B. LAW ENFORCEMENT AND POLICING

In this study, we comprehensively reviewed the fields of AI-based law enforcement and policing (including predictive policing). This is because public safety and law enforcement discussions exhibit parallel trends and often encompass continuous processes.

The most prominent issues in the field of public safety were the potential ethical problems such as AI bias and human rights violations [32-34]. These discussions were frequently raised as issues in the field of remote biometric identification (RBI) systems using AI [35]. AI-based biometric identification technology is most commonly used in border control, such as passport screening for foreigners. However, using AI for border control has also raised ethical concerns [36-38]. To address these concerns, new regulations should be created and AI algorithms should be made transparent and accountable [39].

Biometric identification and predictive policing using AI naturally lead to subsequent law enforcement issues, and this discussion can be divided into optimistic claims that utilizing AI can improve the effectiveness and fairness of court rulings [40-42], and concerns about exacerbating inequality, weakening judicial authority, increasing the possibility of political exploitation, and strengthening digital authoritarianism [42, 43]. This eventually extends to the impact of AI systems on political systems, that is, the impact on democratic systems. Analyses of experimental democratic systems using AI technology [44-46] show the possibility of a new democratic methods using AI technology, but there may also be various side effects such as the problem of false information in the election process [43, 47], and illegal use of personal information [48]. It shows that AI can be an important opportunity for future democracy while also posing a serious threat, which will be a very important research opportunity in the future.

Last but not least, there is a field that is highly underexplored. It is the field of malicious use of AI. This refers to crimes using AI (AIC; Artificial Intelligence Crime) [49], and discussions on the military use of AI [50, 51]. Technologies utilizing AI to attack other individuals or nations are being secretly researched in many countries and are even being used in actual battlefields, but there is a lack of academic research on this topic. Areas such as autonomous weapon systems (AWS) without human intervention, AI-based surveillance systems, and cyber warfare that maliciously utilize AI will occupy a very important domain of future national security. However, academic discussions on related ethical issues, accountability, and international law are relatively lacking. Moreover, even the EU AI Act, known as the world's strongest law, does not address these issues. In the future, academia needs to approach this problem more proactively. To ensure safer utilization of AI technology in future society, it is essential to establish a new normative system for the malicious use of AI.

### C. BIOMETRIC IDENTIFICATION AND CLASSIFICATION

AI-based biometric recognition and classification technologies are advancing human-computer interaction (HCI) by enabling the recognition and analysis of user facial features or emotions. Key applications include assisting individuals with autism in understanding others' emotions [52], enabling customer service to respond based on identified customer emotions [53], and developing affective tutoring systems that adapt to learners' emotions to enhance motivation [54].

However, the use of AI-based biometric recognition raises significant ethical and privacy concerns due to its reliance on sensitive personal data. Insufficient security, transparency, reliability, and explainability in processing sensitive information like emotions and facial features can lead to privacy violations and data misuse [55]. Collecting biometric data in public spaces poses particular challenges, such as difficulties in fulfilling user notification requirements and potential public resistance. The retraction of a facial recognition system initially deployed for security purposes by a Japanese railway company due to privacy concerns and unclear operational policies [56], and the potential for misuse exemplified by the EU's iBorderCtrl system [57], underscore these concerns.

While AI-based biometric recognition technologies are being actively developed across various fields, in-depth discussions regarding their application in public sectors and policy-making remain scarce. This is reflected in the reservations of regulatory experts who oppose the adoption of these technologies due to concerns over privacy violations and ethical issues [58]. Consequently, to facilitate the smooth integration of AI-based biometric systems into the public sphere, efforts are needed to enhance their reliability and public acceptance. In conclusion, resolving the ethical challenges of AI-based biometric systems and ensuring their safe deployment in the public sector urgently requires advancements in explainable AI (XAI) to ensure transparency and accountability, the establishment of robust personal data protection mechanisms, and the development of clear legal regulations on the scope and methods of biometric data utilization.

### D. CRITICAL INFRASTRUCTURE

The EU AI Act categorizes AI systems intended as safety components in critical infrastructure sectors, including digital infrastructure, transportation, energy, and healthcare, as high-risk.

In the digital infrastructure field, there is a strong focus on incorporating AI into the cyber world [59], developing control systems for biometric authentication in infrastructure security [60], implementing generative AI in the service industry [61], and addressing AI security in cloud environments [62]. These topics primarily pertain to the adoption of AI and the corresponding security concerns in the context of digital transformation. The transportation industry has placed particular emphasis on utilizing AI for unmanned automation and conducting research on cyber security. This includes the implementation of AI for monitoring intelligent railway infrastructure [63], the implementation of autonomous shuttle services [64], the utilization of unmanned aviation systems [65], and the adoption of cyber security standards to protect infrastructure [66]. In the energy sector, research has been conducted on the use of AI for transmission network design [67], power quality management [68], and protection of infrastructure against extreme weather conditions [69]. The medical field has focused on researching the implementation of AI in public healthcare [70] and AI-based elderly healthcare assistance [71]. In the public administration sector, issues such as barriers to AI adoption [72], legal framework [73], and information asymmetry [74] have been addressed. In the financial sector, research has explored sentiment analysis using natural language processing to enhance financial services [75].

The study of AI governance within critical infrastructure has received less attention compared to other fields, and has focused more on exploring the potential applications of AI rather than regulation. Particularly, the intelligent railway infrastructure monitoring system in the transportation sector poses a risk of unintended human surveillance, and AI-based customer sentiment analysis in the financial sector can lead to negative consequences in case of misinterpretation. Therefore, further research is necessary to establish a governance-based framework for analyzing potential risks associated with AI use in critical infrastructure, develop AI impact assessment tools, and conduct risk assessment aligned with EU AI Act.

### E. EDUCATION AND TRAINING

The use of AI in education and vocational training has been one of the most active areas of research among the AI systems covered in this study. Notably, higher education [76, 77] and the medical field [78, 79] have shown significant interest in AI technology.

As revealed in the text analysis, in higher education, studies have investigated student perceptions of AI adoption [77] and explored how AI can enhance learning outcomes and improve educational governance [76]. Alongside these investigations, there is a strong interest in utilizing generative AI, particularly ChatGPT. Over 30 studies mentioned ChatGPT or generative AI, with a significant portion focusing on how these technologies can be used to enhance students' self-regulated learning abilities [80, 81]. However, concerns have also been raised regarding the potential for generative AI to facilitate academic misconduct and hinder the development of critical thinking skills [82]. Therefore, further research is needed to analyze the multifaceted effects of AI technology on learning outcomes, motivation, and attitudes.

In the medical field, numerous studies have aimed to improve the efficiency of training involving complex and advanced interpretive skills (e.g., X-ray, CT, and MRI image interpretation) through the use of AI [78, 79]. These studies demonstrate that AI can be effectively used in the training of medical professionals. In contrast, research on AI applications in vocational training and workplace education has been relatively scarce. With few exceptions, such as research on AI-based safety training in manufacturing [83], vocational training was generally mentioned only as a secondary consideration in studies primarily focused on improving industrial process efficiency through AI. Thus, further research exploring the potential of AI in vocational training is needed.

The EU AI Act regulates the use of AI in education, particularly in areas such as admissions, grading, educational assessment, and plagiarism detection. While few of the analyzed papers directly addressed the specific regulatory scope of the EU AI Act in education, many studies indirectly relate to the Act's provisions, covering topics such as enhancing educational outcomes through AI, strengthening self-directed learning, and personalizing education. These areas are indirectly associated with assessments of grades and educational attainment. Therefore, to ensure the effective integration of AI in education, future efforts must focus on securing transparency in assessment processes as stipulated by the EU AI Act, developing specific measures to enhance the reliability of AI systems. Furthermore, it is crucial to develop AI utilization models optimized for educational settings by understanding the perceptions and needs of various stakeholders, including learners, educators, and policymakers.

### F. WORKER MANAGEMENT

AI has been used in managing employees to make decisions like hiring, promoting, and firing automatically [84-86]. However, it is crucial to establish governance protocols to safeguard workplace data [84] as there are potential ethical issues, such as the discrimination. In particular, establishing an audit framework is required to enhance the equity and transparency of AI-based hiring systems [85]. Additionally, concerted efforts and strategies are necessary to mitigate potential human biases throughout the hiring process [86].

Research has demonstrated that AI-based management can lead to negative psychological and ethical consequences for workers [87]. Therefore, it is imperative to prioritize worker consent and rights when implementing AI [88]. In the era of AI, discussions are underway on the direction of changes in labor laws to protect workers' rights [89]. In addition, there are proposals to include obligations for labor groups to participate in the implementation process of AI in the workplace and develop safety standards in labor law [90].

AI can influence both worker productivity and the enhancement of work environments. AI tools can contribute to increased worker productivity and demonstrate the potential for human-AI collaboration [91]. While AI-based human resource management system can enhance organizational performance [92], its acceptance may vary based on the employee's locus of control [93]. Employees who are internally controlled may view the implementation of AI as a potential threat to their employment.

In conclusion, AI-based worker management necessitates discourse on multifaceted aspects, including ethical decision-making, worker-AI interactions, protecting worker rights, and productivity enhancements. Comprehensive research on legal and institutional frameworks is crucial to mitigate ethical and rights-infringement concerns.

## G. ESSENTIAL PRIVATE AND PUBLIC SERVICES

AI research in essential public services centers on integrating public services, enhancing citizen interactions, addressing ethical and reliability concerns in the public sector, and exploring legal and policy implications. AI has the potential to improve the effectiveness of public services [94] and enhance data-driven decision-making [95]. AI-based electronic government systems can strengthen local democracy [96]. In European countries, research has been conducted on the implementation of AI-based public services [97] and the trust and acceptance of citizens [98]. While the introduction of AI can bring about positive impacts, it is crucial to establish institutional measures to ensure ethical use [99] and reliability [44]. Furthermore, it is important to establish legal responsibility and framework for the use of AI [100, 101].

Active research explores the applications of AI in critical private services, including insurance, finance, and healthcare. AI has demonstrated enhanced reliability and customer satisfaction in insurance and financial services [102, 103]. Technological advancements enabled by AI have contributed to improved efficiency within the insurance and finance sectors [104]. In healthcare, AI holds significant promise for service optimization [105], patient flow management [106], and advancements in disease prevention and treatment [107].

The utilization of AI technology is rapidly increasing in both the public and private sectors; however, it is imperative to address critical concerns such as reliability, ethics, and data protection. For electronic government and AI-based public services, a policy-based foundation is crucial to enhance trust and citizen engagement. Ignoring security and ethical issues could hinder the private sector's long-term growth. In short, while AI has the potential to improve both public and private services, it's essential to balance technology and human interaction, enforce ethical and legal regulations, and encourage collaboration and complementary development strategies between the public and private sectors.

## VI. CONCLUSION

This study employed bibliometric analysis and text mining techniques to comprehensively analyze recent research trends in AI governance, specifically within the context of the emerging EU AI Act framework. The analysis reveals that research on AI governance and regulation constitutes a relatively small portion of the overall AI research landscape. Notably, studies focusing on policy aspects related to AI systems regulated under the EU AI Act are even more limited. This indicates that the development of relevant policies and regulations is lagging behind the rapid pace of AI technological advancement. Despite this, the study confirms that AI governance research is expanding across various academic disciplines, with ethics, privacy, and generative AI-related topics emerging as major areas of interest. Furthermore, the prominence of multidisciplinary journals and open-access publications in AI governance research underscores the importance of rapid dissemination and sharing of findings in this nascent field. The analysis also reaffirms that while the US and China are leading in AI governance research, variations exist in research themes and approaches, influenced by national contexts, cultural factors, and specific policy priorities. This study holds significance as it systematically analyzes AI governance research trends based on the concrete regulatory framework of the EU AI Act and provides a comprehensive overview of major research themes, keywords, and national research trends.

Despite its contributions, this study has several limitations that simultaneously highlight key areas for future research. First, the scarcity of research on Type C systems (social scoring) precluded an in-depth analysis of this specific system. This may be attributed to the limited scope of the definitions of "social behavior" and "personal characteristics" within the EU AI Act. Future research should employ more comprehensive search terms to capture a wider range of relevant studies on this topic. Second, the reliance on the Web of Science database means that studies indexed in other databases, such as Scopus and ArXiv, were not included. Future research should utilize multiple databases to provide a more holistic view of research trends. This is particularly important in the rapidly evolving field of AI, where many studies are shared as preprints rather than through traditional peer-reviewed outlets [5]. The use of the WOS database might have limited the ability of this study to capture the most recent trends in AI research. Building on the implications of this study, future research should leverage preprint databases like ArXiv to analyze the latest developments in AI and contribute to practical policy-making.

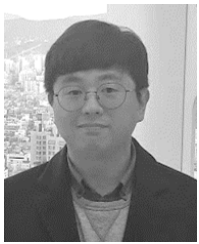

**Byeong-Je Kim** is a research assistant professor at the Institute of Social Data Science, Pohang University of Science and Technology (POSTECH), Republic of Korea. His research interests focus on disaster management policy, risk perception, and safety education.

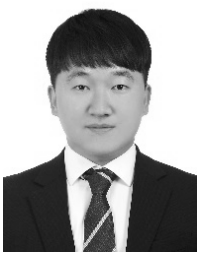

**Seunghoo Jeong** is a senior researcher at the R&D Planning and Strategy Division, Korea Railroad Research Institute (KRRI), Republic of Korea. His current research interests include structural health monitoring of critical infrastructure (e.g., railway system), risk perception, and disaster management.

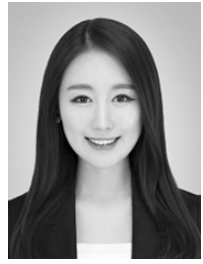

**Bong-Kyung Cho** is the team leader of the Strategy Group at INTERX, an AI company. Her research interests focus on AI regulation, Carbon regulation, and governance.

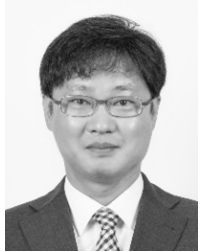

**Ji-Bum Chung** is an associate professor at the Department of Civil, Urban, Earth, and Environmental Engineering, Ulsan National Institute of Science and Technology (UNIST), Republic of Korea. His research interests focus on risk communication/management and disaster management.